\documentclass[aps,prb,twocolumn,superscriptaddress]{revtex4-1}
\usepackage{bm}          
\usepackage{amsfonts}   
\usepackage{amsmath,amssymb,bbm} 
\usepackage{graphicx}    
\usepackage{xspace}		

\newcommand{\luvo}{Lu$_2$V$_2$O$_7$\xspace}

\begin{document}

\title{{\it ab-initio} Approaches for Low-Energy Spin Hamiltonians}
\date{\today}

\begin{abstract}
Implicit in the study of magnetic materials is the concept of spin Hamiltonians, which emerge as the low-energy theories of correlation-driven insulators. In order to predict and establish such Hamiltonians for real materials, a variety of first principles {\it ab-initio} methods have been developed, based on density functional theory and wavefunction methodologies. In this review, we provide a basic introduction to such methods and the essential concepts of low-energy Hamiltonians, with a focus on their practical capabilities and limitations. We further discuss our recent efforts toward understanding a variety of complex magnetic systems that present unique challenges from the perspective of {\it ab-initio} approaches.
\end{abstract}

\author{Kira Riedl}
\affiliation{Institut f\"ur Theoretische Physik, Goethe-Universit\"at Frankfurt,
Max-von-Laue-Strasse 1, 60438 Frankfurt am Main, Germany}
\author{Ying Li}
\affiliation{Institut f\"ur Theoretische Physik, Goethe-Universit\"at Frankfurt,
Max-von-Laue-Strasse 1, 60438 Frankfurt am Main, Germany}
\author{Roser Valent{\'\i}}
\affiliation{Institut f\"ur Theoretische Physik, Goethe-Universit\"at Frankfurt,
Max-von-Laue-Strasse 1, 60438 Frankfurt am Main, Germany}
\author{Stephen M. Winter}
\email{winter@physik.uni-frankfurt.de}
\affiliation{Institut f\"ur Theoretische Physik, Goethe-Universit\"at Frankfurt,
Max-von-Laue-Strasse 1, 60438 Frankfurt am Main, Germany}

\maketitle


\section{Introduction} At the heart of the study of quantum spin materials is
the concept of low-energy (spin) Hamiltonians, which describe the magnetic
states relevant at experimental energy
scales~\cite{patrik1999lecture,mattis2012theory,fazekas1974ground,illas2000magnetic,de2006unified,bencini2008some,xiang2013magnetic}.
The emergence of such spin degrees of freedom occurs due to the localization of
unpaired electrons in solids by the effects of mutual Coulomb repulsion, thus
forming a Mott insulating state~\cite{mott1968metal,imada1998metal}. In such cases, effective
spin degrees of freedom provide an efficient method of describing the
low-energy
states~\cite{macdonald1988t,moriya1960anisotropic,yildrim1995anisotropic}. 
Actually, they represent far more complex electronic states with specific details
of charge, orbital and lattice degrees of freedom being embedded in the
coupling constants of effective
Hamiltonians~\cite{tokura2000orbital,khaliullin2005orbital,gardner2010magnetic,witczak2014correlated,rau2016spin}.
Formally, the most general interactions can be expanded as products of spin
operators (or equivalently Stevens operators~\cite{stevens1952matrix})
representing the local degrees of freedom at each magnetic site,
\begin{align}
 \mathcal{H}_{\rm spin} = & \ \sum_{i,\mu} a_i^\mu S_i^\mu +
\sum_{ij,\mu\nu} b_{ij}^{\mu\nu} S_i^\mu S_j^\nu \nonumber \\ & \ \hspace{10mm}
+ \sum_{ijk,\mu\nu\xi}c_{ijk}^{\mu\nu\xi}S_i^\mu S_j^\nu S_k^\xi + ...
\end{align}
with $\mu,\nu,\xi \in \{x,y,z\}$. The couplings constants
 $a_i^\mu, b_{ij}^{\mu\nu}, ...$  can
include all terms respecting the symmetry of the lattice and the structure of
the quantised Hilbert space. Therefore, provided some ingenuity in materials
design, a wide variety of such Hamiltonians can be realized in real materials.
The appeal of such materials is the fact that the local spins represent the
simplest of local quantum variables, allowing intriguing connections to simple
models of statistical physics and quantum information. Thus, an incredibly rich
variety of physical states and phase transitions can, in principle, be realized
in quantum spin materials, from highly entangled spin
liquids~\cite{balents2010spin,gingras2014quantum,zhou2017quantum} to classical
and quantum critical phenomena~\cite{sachdev2000quantum,sachdev2011quantum}.

 The ability
to predict and evaluate low energy Hamiltonians of specific materials
constitutes a vital contribution to the development and understanding of
complex spin systems. In this pursuit, various {\it ab-initio}
methods have been developed to provide first-principles estimates of the
coupling constants $a_i^\mu, b_{ij}^{\mu\nu}, ...$, based on different
approximations. Through the use of these methods, insight can be
gained into both, the coupling constants describing known materials, and the potential for tuning such coupling
constants by chemical or physical means. 

The purpose of this short review is to
motivate the basic concepts in the mapping of electronic Hamiltonians to
effective spin models.  We discuss some popular {\it ab-initio} methods, with a
focus on relative merits and current challenges in improvement of the method.
Finally, we discuss some applications towards quantum spin materials of recent
interest, highlighting the contributions from different {\it ab-initio} methods
towards the understanding of the underlying spin Hamiltonians. This review is
not intended to present a complete picture of the field, but rather to provide
some perspective, particularly for the non-expert.


\section{Low-Energy Hamiltonian Concept} \label{sec:concept}

\begin{figure}
\includegraphics[width=0.85\columnwidth]{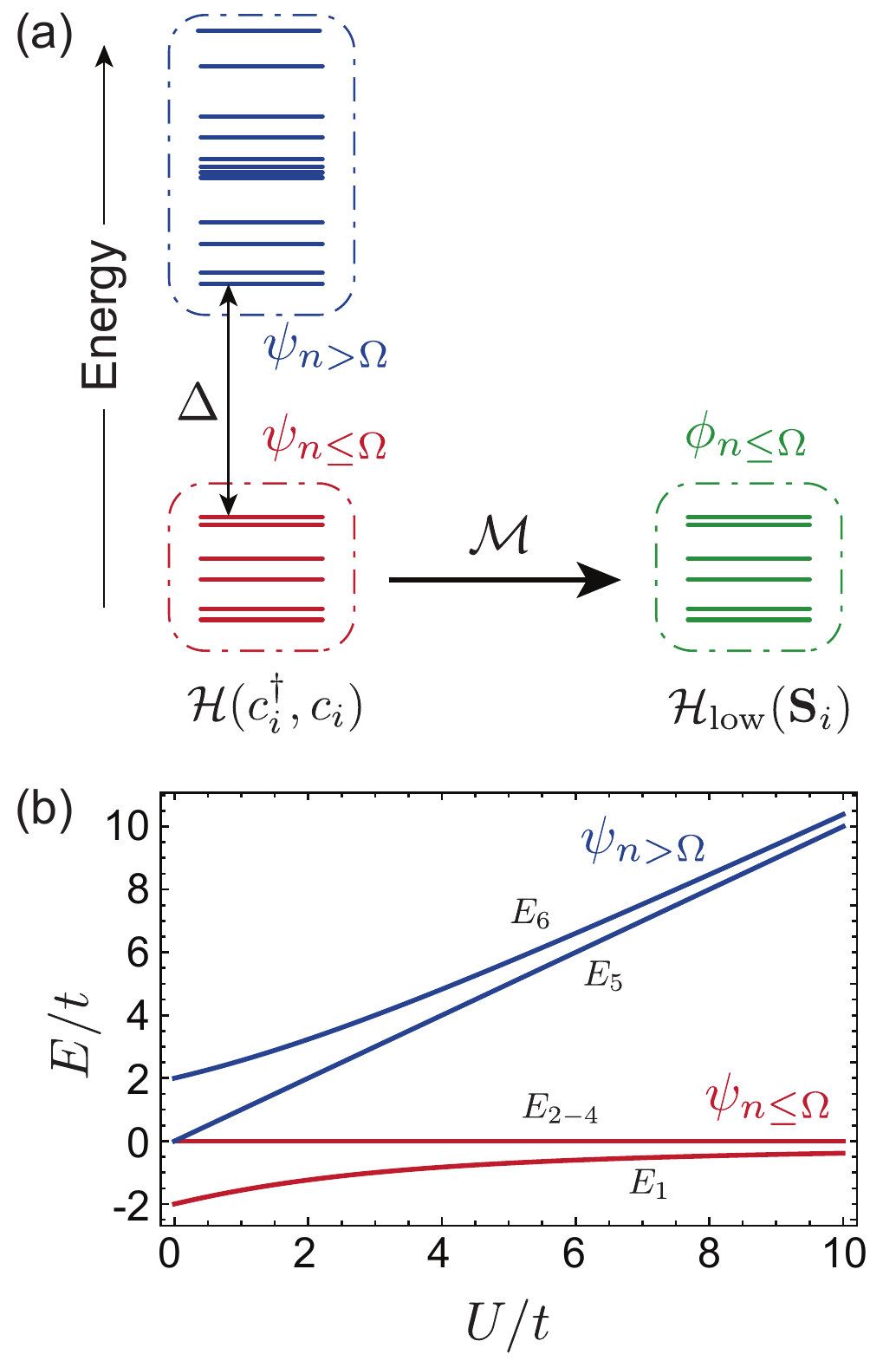}
\caption{(a) Schematic construction of low-energy Hamiltonian $\mathcal{H}_{\rm low}$ via the mapping $\mathcal{M}$. (b) Spectrum of the simple 2-site Hubbard model, divided into low and high energy states.}
\label{fig:lowenergy}
\end{figure}

In order to briefly introduce the concept of low-energy theories, we may consider a Hamiltonian $\mathcal{H}$, with an associated Hilbert space $\mathcal{S}$ that is spanned by the eigenstates $\{|\psi_n\rangle\}$. Here, the quantum numbers $n = 1,2, \  ... ,\  N$ label the eigenstates according to increasing energy $E_n = \langle \psi_n|\mathcal{H}|\psi_n\rangle$, such that $E_n \leq E_{n+1}$. For our purpose, $\mathcal{H}$ could represent an electronic Hamiltonian for a solid, with a large Hilbert space consisting of a variety of lattice, charge, spin, and orbital degrees of freedom. In many cases, we will find that the spectrum of $\mathcal{H}$ contains a certain number of the lowest energy eigenstates (with $n \leq \Omega < N$) that are widely separated from higher energy states, as illustrated in Fig.~\ref{fig:lowenergy}(a). The origin of the energy gap $\Delta$ may be, for example, the effects of crystal field or Coulomb interactions, which select low-energy states with similar charge, lattice, and orbital configurations. Generally, those states with $n>\Omega$ will contribute very little to the experimental response at low temperatures and frequencies ($T,\omega \ll \Delta$), indicating both a conceptual and computational advantage to ``integrate out'' the higher energy states. In the case where the number of states below the gap follows $\Omega = (2S+1)^{n_{\rm sites}}$, we may describe the low-energy response in terms of effective spins $S$ living at sites of number $n_{\rm sites}$. In order to do so, we define a new low-energy Hamiltonian $\mathcal{H}_{\rm low}$, and a new low-energy Hilbert space $\mathcal{S}_{\rm low}$ spanned by the eigenstates $\{|\phi_n\rangle\}$, with $n = 1,2, ..., \Omega$. We require only that this new Hamiltonian reproduces exactly the spectrum of the original Hamiltonian $\mathcal{H}$, such that $\langle \phi_n | \mathcal{H}_{\rm low}| \phi_n\rangle = \langle \psi_n | \mathcal{H} |\psi_n\rangle$. This sole requirement preserves the freedom to write $\mathcal{H}_{\rm low}$ and $\mathcal{S}_{\rm low}$ in terms of {\it any} convenient basis or variables, although we will exclusively focus on spin Hamiltonians here. 

For concreteness, let us demonstrate the concept on a two-site Hubbard model
at half-filling, with a single orbital per site, as described by the electronic
Hamiltonian $\mathcal{H} = \mathcal{H}_t + \mathcal{H}_U$ with:
\begin{align}
\mathcal{H}_t = & \  -t \sum_{\sigma=\{\uparrow,\downarrow\}} (c_{1,\sigma}^\dagger c_{2,\sigma} + c_{2,\sigma}^\dagger c_{1,\sigma}) \\ 
\mathcal{H}_U = & \ U(n_{1,\uparrow}n_{1,\downarrow} + n_{2,\uparrow}n_{2,\downarrow})
\end{align}
where $c_{i,\sigma}^\dagger$ creates an electron at site $i$, with spin $\sigma \in \{\uparrow,\downarrow\}$. 
The Hilbert space $\mathcal{S}$ includes six states. The lowest energy eigenstate is a spin singlet $(S = 0)$ for any finite $t/U$, given by:
\begin{align}
|\psi_1\rangle = & \ \gamma \left( |\uparrow_1 \downarrow_2\rangle +|\uparrow_2\downarrow_1\rangle \right) \nonumber \\ & \ \hspace{5mm}+ \sqrt{1-\gamma^2} \left( |\uparrow_1\downarrow_1\rangle + |\uparrow_2\downarrow_2\rangle \right)\\
E_1 = & \ \frac{1}{2}\left( U-\sqrt{16t^2 + U^2}\right) \\
\gamma = & \ \frac{1}{\sqrt{2}}\frac{2t}{\sqrt{E_1^2 + (2t)^2}}
\end{align}
This is followed by a 3-fold degenerate spin triplet ($S=1$), given by $\left|\psi_2\right\rangle =\left|\uparrow_1\uparrow_2\right\rangle$, $\left|\psi_3\right\rangle =  \frac{1}{\sqrt{2}}(\left|\uparrow_1\downarrow_2\right\rangle - \left|\uparrow_2\downarrow_1\right\rangle)$, and $\left|\psi_4\right\rangle = \left|\downarrow_1\downarrow_2\right\rangle$, with equal energies $E_{2-4}  = 0$.

Finally, the last two states are spin singlets, given by $|\psi_5\rangle = \frac{1}{\sqrt{2}}\left( |\uparrow_1\downarrow_1\rangle - |\uparrow_2\downarrow_2\rangle \right)$ and $|\psi_6\rangle =  \gamma ( \left|\uparrow_1\downarrow_1\right\rangle + \left|\uparrow_2\downarrow_2\right\rangle )- \sqrt{1-\gamma^2} ( \left|\uparrow_1 \downarrow_2\right\rangle +\left|\uparrow_2\downarrow_1\right\rangle)$, with energies $E_5 = U$ and $E_6 = \frac{1}{2}\left( U+\sqrt{16t^2 + U^2}\right)$, respectively. The evolution of these state energies with $U/t$ is shown in Fig.~\ref{fig:lowenergy}(b).

In the limit $U\gg t$, the energies of the first four states are of order $\sim 0$, while the latter two states have energies of order $U$. In order to describe the low-energy response in this limit, it is therefore advantageous to consider an effective Hamiltonian that treats only the lowest four states explicitly. To do so, we first choose a low-energy Hilbert space $H_{\rm low}$ that is spanned by the pure spin states $\{|\uparrow_1\uparrow_2\rangle, |\uparrow_1\downarrow_2\rangle, |\uparrow_2\downarrow_1\rangle, |\downarrow_1\downarrow_2\rangle\}$. While these states become exact eigenstates of $\mathcal{H}$ in the limit $t/U \to 0$, such a condition is not generally necessary. We then associate states in $\mathcal{S}_{\rm low}$ with the states in $\mathcal{S}$ through a projective mapping $\mathcal{M}: \mathcal{S} \to \mathcal{S}_{\rm low}$, such that $|\phi_n\rangle = \mathcal{M}(|\psi_n\rangle)$. Under such a mapping, the states transform as:
\begin{align}
 |\phi_n\rangle =  \left\{ \begin{array}{cc}  \frac{1}{\sqrt{2}}\left( |\uparrow_1 \downarrow_2\rangle +|\uparrow_2\downarrow_1\rangle \right)& n= 1 \\ |\psi_n\rangle & n=2,3,4 \\ 0 & n=5,6\end{array}\right.
%
%
\end{align}
Since $|\psi_{2-4}\rangle$ are already pure spin states, they are unaffected by the mapping. In terms of the new states, the condition $\langle \psi_n | \mathcal{H} | \psi_n\rangle = \langle \phi_n | \mathcal{H}_{\rm low} | \phi_n \rangle$ then defines the low-energy effective Hamiltonian $\mathcal{H}_{\rm low} = \mathcal{M}(\mathcal{H})$, which is an isotropic Heisenberg Hamiltonian:
\begin{align}
\mathcal{H}_{\rm low} = & \ J \left(\mathbf{S}_1 \cdot \mathbf{S}_2 - \frac{1}{4} \right)
\end{align}
where the coupling constant is a function of $t$ and $U$:
\begin{align}
J =& \  -E_1 = \frac{1}{2}\left( \sqrt{(4t)^2 +U^2}-U\right)
\end{align}
The physical origin of this coupling constant is that the true singlet ground state $|\psi_1\rangle$ has a larger variational flexibility than the triplet states $|\psi_{2-4}\rangle$, and may obtain a lower kinetic energy through mixing of charge neutral and charge separated states. Based on this simple example, we may highlight several key aspects of low-energy effective spin Hamiltonians:

(i) The form of the low-energy Hamiltonian $\mathcal{H}_{\rm low}$ depends on {\it both} the low-energy spectrum of $\mathcal{H}$, {\it and} the specific choice of mapping $\mathcal{M}$. For this reason, a low-energy Hamiltonian is only uniquely defined up to the mapping $\mathcal{M}$, which relates the effective low-energy spin degrees of freedom to the real physical states. In this simple example, full knowledge of the eigenstates of $\mathcal{H}$ has allowed us to employ an ``optimal'' mapping, which maximizes $\langle \psi_n|\phi_n\rangle$, and preserves all symmetries. As a result, $\mathcal{H}_{\rm low}$ explicitly displays all the symmetries of $\mathcal{H}$ such as SU(2) spin-rotational invariance, and the exact eigenstates of $\mathcal{H}$ approaching the low energy states (i.e. $|\psi_n\rangle \to |\phi_n\rangle$) in the limit $t/U \to 0$. These features are not required for a valid low-energy theory, but nonetheless help us to intuitively understand the meaning of the coupling constants and low-energy degrees of freedom. As discussed in more detail in Sec.~\ref{sec:methods}, it may not always be possible to specify the mapping $\mathcal{M}$ assumed by an {\it ab-initio} method. This leads to some ambiguity in the computed coupling constants, which must be considered when comparing different methods.

(ii) It is further important to remember that the low-energy spin degrees of freedom only represent effective variables, such that operators acting in the full Hilbert space $\mathcal{S}$ must be mapped into $\mathcal{S}_{\rm low}$, or else they may yield incorrect expectation values. That is, in general:
\begin{align}
\langle \psi_n| \hat{O} | \psi_n\rangle = \langle \phi_n|\mathcal{M}( \hat{O}) | \phi_n\rangle \neq \langle \phi_n| \hat{O} | \phi_n\rangle
\end{align}
To illustrate this, we can consider the action of $S_1^2$, which measures the spin multiplicity at site 1. In terms of the electronic operators, this is:
\begin{align}
\hat{S}_1^2 =\frac{3}{4}\left(n_{1,\uparrow}+n_{1,\downarrow} - 2n_{1,\uparrow}n_{1,\downarrow} \right)
\end{align}
If we were to measure this operator on the ground state of $\mathcal{H}$, it would yield $\langle \psi_1| \hat{S}_1^2 |\psi_1\rangle = (3\gamma^2/4)$, which differs from the naive action of the operator on the pure spin states in $\mathcal{S}_{\rm low}$, for which $\langle \phi_1|\hat{S}_1^2 | \phi_1\rangle = 3/4$. In fact, the correct expectation values are obtained only by projecting $\hat{S}_1^2$ into the low-energy space, to yield:
\begin{align}
\mathcal{M}(\hat{S}_1^2) =\frac{3}{4} +\frac{3}{4}\left(\frac{J}{\sqrt{(4t)^2+U^2}} \right) \left(\mathbf{S}_1 \cdot \mathbf{S}_2 - \frac{1}{4} \right)
\end{align}
Interestingly, the local operator in the full Hilbert space becomes non-local in terms of the fictitious low-energy spin states. The reduction of $\langle \hat{S}_1^2\rangle$ is due to the formation of a covalent bond as $U/t$ is reduced.


\section{Numerical and Analytical Methods} \label{sec:methods}

In this section, we review some of the methods employed in the estimation of coupling constants in low-energy spin Hamiltonians, with a specific focus on their relative merits and deficiencies. These methods can be divided into two categories: (i) those which are fully (or nearly fully) {\it ab-initio} such as Density Functional Theory (DFT) and Multi-Reference ``Quantum Chemistry'' methods, and (ii) semi-{\it ab-initio} methods based on approximate electronic Hamiltonians such as perturbation theory and ``hybrid'' cluster diagonalization methods. This list is not complete, but includes the most commonly used approaches for studying magnetic insulators.\\


\subsection{Perturbation Theory} \label{sec:pert}

In perturbative approaches, approximate expressions for the coupling constants are obtained by expanding about a well-understood limit of a model electronic Hamiltonian (usually a Hubbard-like model incorporating the relevant orbitals at each magnetic site). This method may yield analytical forms for all symmetry-allowed spin coupling constants in terms of the parameters of the model electronic Hamiltonian. As such, perturbation theory does not represent an {\it ab-initio} method by itself.

Consider a Hamiltonian divided as $\mathcal{H} = \mathcal{H}_0 + \lambda \mathcal{H}_1$. An effective Hamiltonian can be developed using Brillouin-Wigner perturbation theory. We choose to label states in our low-energy Hilbert space $\mathcal{S}_{\rm low}$ according to eigenstates of unperturbed Hamtiltonian $\mathcal{H}_0$ with $n\leq \Omega$, for which $E_n^0 = \langle \phi_n|\mathcal{H}_0 | \phi_n\rangle$. We then define a projection operator onto the low energy states of $\mathcal{H}_0$ as:
\begin{align}
\mathcal{P} = \sum_{n \leq \Omega} |\phi_n\rangle \langle \phi_n|
\end{align}
with $\mathcal{Q} = 1-\mathcal{P}$ giving the projection onto the high energy states of $\mathcal{H}_0$. In terms of such operators, the eigenstates of the full Hamiltonian are then given by:
\begin{align}
| \psi_n \rangle =& \  \left( 1-\lambda \frac{1}{E_n - \mathcal{H}_0}\mathcal{Q} \mathcal{H}_1\right)^{-1} |\phi_n\rangle \\
= & \ |\phi_n\rangle + \lambda \sum_{m>\Omega} |\phi_m\rangle \frac{\langle \phi_m | \mathcal{H}_1 | \phi_n\rangle}{E_n - E_m^0} + ...
\end{align}
where $E_n = \frac{\langle \psi_n |\mathcal{H}|\psi_n\rangle}{\langle \psi_n|\psi_n\rangle}$ is the exact energy of the given state. Note that the states $\{|\psi_n\rangle\}$ defined in this way are not strictly normalized, but rather follow the ``intermediate normalization condition'' where $\langle \psi_n | \phi_n \rangle = 1$. The effective Hamiltonian is then formally:
\begin{align}\nonumber
\mathcal{H}_{\rm low} =& \  \mathcal{P} \left[\mathcal{H}_0 + \lambda \mathcal{H}_1\left(1-\lambda\frac{1}{E_n-\mathcal{H}_0}\mathcal{Q} \mathcal{H}_1\right)^{-1} \right]\mathcal{P} \\
 = & \ \left(\mathcal{H}_0 + \lambda \mathcal{H}_1\right)+ \lambda^2\sum_{m>\Omega} \frac{\mathcal{H}_1 |\phi_m\rangle \langle \phi_m| \mathcal{H}_1}{E_n - E_m^0} + ...
\end{align}
Carrying out this procedure for the example of a 2-site Hubbard model of Sec.~\ref{sec:concept}, we may choose $\mathcal{H}_0 = \mathcal{H}_U$ and $\mathcal{H}_1 = \mathcal{H}_t$, and expand in powers of $t/U$. Up to second order in $t$, this yields the familiar result:
\begin{align}
\mathcal{H}_{\rm low} = J \left(\mathbf{S}_1 \cdot \mathbf{S}_2 - \frac{1}{4} \right); \ \ J = \frac{4t^2}{U} + \mathcal{O}\left(\frac{t^4}{U^3}\right)\label{eq:pert}
\end{align}
A significant advantage of perturbation theory is that it represents a well-defined approximation scheme for {\it both} $\mathcal{H}_{\rm low}$ and the mapping $\mathcal{M}$, which allows approximate expressions for any expectation value to be derived within the same scheme. However, perturbative expressions may prove to be misleading if important higher order contributions are neglected. For example, obtaining expressions for longer-range couplings beyond nearest neighbour would usually require higher orders to be computed, which may quickly become unwieldy. As a result, non-perturbative, fully {\it ab-initio} methods are desirable, such as broken symmetry DFT or quantum chemistry cluster methods.  \\


\subsection{Total Energy (Broken Symmetry) DFT} One of the most widely used methods for estimating spin exchange constants in based on density functional theory (DFT) approaches~\cite{noodleman1981valence,noodleman1986ligand,noodleman1995orbital,yamaguchi1988ab,yamaguchi1989antiferromagnetic,ruiz1999broken,tsirlin2009extension,lebernegg2013magnetism,glasbrenner2015effect}. These approaches are relatively inexpensive computationally, which allows for rather large systems to be treated without further approximations. For example, the full environment of a periodic crystal may be included. In many cases, this feature allows second and third neighbour couplings to be investigated without prohibitive computational cost. This feature can be particularly advantageous for studying some strongly frustrated magnetic systems, where such couplings may ultimately select the ground state. 

\begin{figure}[t]%
\includegraphics*[width=0.95\linewidth]{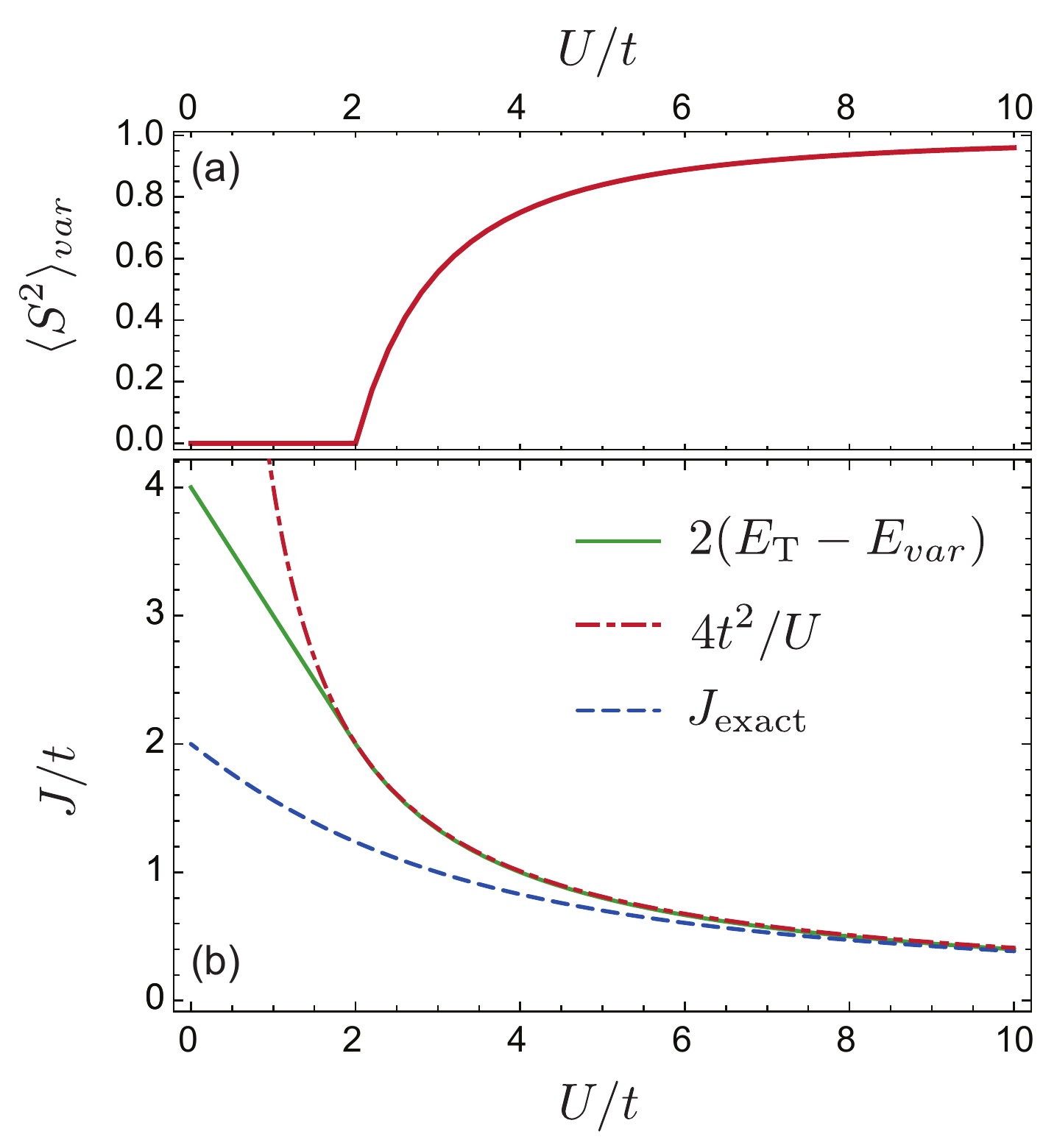}
\caption{Broken symmetry approach for the two-site Hubbard Model. (a) Evolution of the $\langle S^2\rangle$ expectation value for the variational single-determinant broken symmetry state $|\psi_{var}\rangle$ with $S_z=0$. (b) Comparison of the exact Heisenberg coupling constant $J$ to values estimated via Eq.~(\ref{eq:BSS}) and perturbation theory Eq.~(\ref{eq:pert}). \label{fig:BS}}
\label{fig:bss}
\end{figure}

The foundation of DFT is that the ground state energy of an interacting many-body system can be written as a functional of the electronic (spin) density~\cite{PhysRev.136.B864,eschrig2003fundamentals,martin2004electronic,rev_qc_emb_carter_08}. This density can be obtained, in principle, by solving an auxiliary system of independent ``Kohn-Sham'' particles experiencing an effective field that is determined self-consistently~\cite{PhysRev.140.A1133}. In practice, the effective field includes the mean Coulomb potentials, together with an approximate ``exchange-correlation'' potential $V_{xc}$, which is meant to correct for many-body correlations that would appear in the true interacting wavefunction, but are not explicitly captured by the fictitious Kohn-Sham wavefunction. 
Within this approach, the energies of different spin configurations are approximated by performing multiple DFT calculations, constraining the auxiliary Kohn-Sham wavefunctions in each to converge to different spin densities. Since the auxiliary reference states are single-determinant wavefunctions, the energies obtained in this way are typically interpreted as ``classical'' spin energies. The theoretical basis and limitations of this approach have been discussed in detail by several authors~\cite{illas2000magnetic,caballol1997remarks,illas2004extent,ciofini2003dft,neese2004definition,rudra2006accurate}.

To illustrate how BS-DFT is usually applied, let us return to the problem of two-sites introduced in Sec.~\ref{sec:concept}. For the Heisenberg Hamiltonian, the coupling constant $J$ may be obtained from the energy difference between two classical spin configurations with $\langle S_z \rangle = 1$ and 0:
\begin{align}
E_{\rm T} =& \  \langle \uparrow_1\uparrow_2| J \ (\mathbf{S}_1 \cdot \mathbf{S}_2-1/4) | \uparrow_1 \uparrow_2\rangle = 0\\
E_{\rm BSS} = & \ \langle \uparrow_1\downarrow_2| J \ (\mathbf{S}_1 \cdot \mathbf{S}_2 -1/4)| \uparrow_1 \downarrow_2\rangle = -J/2 
\end{align}
which leads to:
\begin{align}
J = & \ 2(E_{\rm T} - E_{\rm BSS}) \label{eq:BSS}
\end{align}
Here, we have introduced the ``broken symmetry singlet'' state $|\uparrow_1 \downarrow_2\rangle$ of Noodleman~\cite{noodleman1981valence,noodleman1986ligand}, which is a classical spin state with $\langle S_z\rangle = 0$. It is not an eigenstate of the quantum Heisenberg Hamiltonian and lacks the full symmetry of the true singlet ground state, which has $E_1 = -J$. However, the fact that $E_{\rm BSS}$ lies half-way between the singlet and triplet energies allows for estimates of $J$, in principle. This suggests, for real materials with multiple magnetic sites, that the various different exchange couplings can be estimated by least squares fitting of the converged energies of different magnetic configurations using expressions analogous to Eq.~(\ref{eq:BSS})~\cite{xiang2013magnetic}.

Since Eq.~(\ref{eq:BSS}) holds for the Heisenberg spin Hamiltonian, let us see if it holds for an electronic Hamiltonian, by comparing the lowest variation energy for single-determinant states with $\langle S_z \rangle = 1$ and 0.
For the hypothetical two-site Hubbard model, the single-determinant state $|\uparrow_1\uparrow_2\rangle$ is an eigenstate for all $U/t$, with energy $E_{\rm T} = 0$. A general single-determinant state with $\langle S_z\rangle = 0$ takes the form:
\begin{align}
|\psi_{var}^{0}\rangle = & \ a_1b_1 |\uparrow_1\downarrow_1\rangle + a_1b_2|\uparrow_1 \downarrow_2\rangle \\ \nonumber & \ + a_2b_1 |\uparrow_2\downarrow_1\rangle + a_2b_2|\uparrow_2\downarrow_2\rangle
\end{align}
with the constraints $a_1^2+a_2^2 = 1$ and $b_1^2+b_2^2 = 1$. Fig.~\ref{fig:BS}
shows the the exchange constant estimated from the lowest variational energy
via $J = 2(E_{\rm T} - E_{var})$. In the limit $U\to \infty$, the lowest
variational energy is indeed obtained for the pure spin wavefunction $
|\uparrow_1\downarrow_2\rangle$. As a result, $E_{var}$ converges to $-J/2$,
which can be correctly interpreted as the classical spin exchange energy. In
fact, for $U/t > 2$, the variational result follows the second order
perturbation theory expression. However, in the opposite limit $U\ll t$, the
variational wavefunction actually converges to the true singlet ground state of
the two-site Hubbard model with $\langle S^2\rangle = 0$, leading to an
overestimation of $J$ by as much as a factor of 2. In practice, this
observation implies potential failure of Eq.~\eqref{eq:BSS} to describe $J$ in terms of total
DFT energies if the converged magnetic moments per magnetic site vary
significantly between different spin configurations~\cite{soda2000ab}. 

At this point it should be re-emphasized that the electronic spin density of the broken symmetry states will almost always differ from the true lowest energy many-body states with equivalent quantum numbers. Paradoxically, the ability to converge the DFT calculations to broken symmetry states with $\langle S^2\rangle >0$ actually relies on (i) the use of approximate exchange-correlation functionals that do not adequately recover the static correlation energy and/or (ii) additional constraints that may impact the accuracies of the obtained coupling constants. A DFT calculation employing the {\it exact} correlation functional with the constraint $\langle S_z \rangle = 0$ would lead to a Kohn-Sham state with $\langle S^2\rangle =0$, and yield the exact energy of the singlet ground state, independent of the broken symmetry starting point~\cite{rudra2006accurate}. In order to remedy this, we might consider instead constraining $\langle S^2\rangle$, in order to ensure convergence to a given spin configuration such as $|\uparrow_1\downarrow_2\rangle$. However, the electronic energy of such a state in the Hubbard approximation would be $\langle \uparrow_1 \downarrow_2| \mathcal{H}_U + \mathcal{H}_t|\uparrow_1\downarrow_2\rangle = 0 = E_{\rm T}$. Thus, imposing such a constraint would completely eliminate the ability to estimate $J$; in the broken symmetry approach, antiferromagnetic contributions arise only from the additional variational freedom of low-spin states. These issues underly the fact that the precise mapping between the auxiliary Kohn-Sham states and the true many-body wavefunctions whose energies they represent is generally unknown. Nonetheless, a wide variety of studies employing broken symmetry DFT have demonstrated fair agreement with experimental estimates of exchange constants~\cite{ruiz1999broken,illas2004extent,martin1997antiferromagnetic,adamo1999calculation}. 
Some general considerations are as follows:

(i) Results obtained with DFT+U or hybrid
functionals~\cite{martin1997antiferromagnetic,jeschke2011multistep,bandeira2012calculation,Jeschke2013,tutsch2014evidence} are far
more adequate than pure LDA or GGA functionals, which do not sufficiently
localize the relevant magnetic orbitals, and may overestimate $|J|$ by several
times. 

(ii)  In general, the computational expense and reliability is adversely affected by the complexity of the spin Hamiltonian. Apart from the intrinsic approximations, the most significant sources of error arise from poor convergence of spin configurations far away from the ground state. 
The number of required spin configurations scales linearly with the number of coupling constants, which increases the likelihood of some poorly convergent configurations adversely affecting the results. This makes extracting {\it all} symmetry-allowed couplings sometimes impossible from DFT calculations. 

(iii) Exchange constants are typically more reliable for higher symmetry cases, e.g. Heisenberg couplings without spin-orbit coupling. Reliable results for anisotropic couplings that arise from spin-orbit coupling (such as the Dzyaloshinskii-Moriya interaction $\mathbf{D}_{ij} \cdot (\mathbf{S}_i \times \mathbf{S}_j)$) and higher order ring-exchange can sometimes be obtained by DFT formulations with non-collinear moments \cite{kubler1988density,PhysRevB.62.11556,takeda2006,xiang2011predicting,glasbrenner2014first,rousochatzakis2015frustration}. However, such calculations require additional care and are not always reliable \cite{PhysRevB.82.220402,PhysRevB.92.054428,0953-8984-30-27-275802}. 

(iv) Since broken symmetry DFT correctly incorporates the essential physics of the isotropic magnetic couplings, it will typically reproduce trends in $J$ (e.g. as a function of crystal structure), even when absolute values prove unreliable. As a result, broken symmetry DFT is often most suitable for tracking differences between structurally similar materials.\\


\subsection{Quantum Chemistry Cluster Methods}

A promising alternative to DFT can be found in wavefunction-based quantum chemistry methods, which explicitly treat the multi-determinant character of the many-body eigenstates~\cite{rev_qc_hozoi_11,rev_qc_emb_paulus_12}. In such methods, an active space of relevant orbitals and corresponding electronic configurations is chosen, within which all many-body effects are explicitly included~\cite{szabo2012modern}. Dynamical screening of the Coulomb interactions are then recovered by including a finite list of particle-hole excitations out of the active space. In the minimal case, the active space would include all states desired to construct the low-energy spin Hamiltonian \cite{calzado2010extending}, while the space of particle-hole excitations would be much larger. As emphasized particularly by Malrieu and coworkers~\cite{calzado2002analysis1,calzado2002analysis2,calzado2009analysis}, the main challenge in applying such methods for calculation of magnetic couplings is selecting a comprehensive list of important particle-hole excitations without incurring a prohibitively large computational cost or overstabilizing particular configurations due to unbalanced truncation. Experience has shown the so-called difference-dedicated configuration interaction (DDCI) scheme~\cite{miralles1992variational,miralles1993specific} to provide very accurate isotropic exchange constants when compared with experimental values or exact Full Configuration Interaction (FCI) \cite{de2006unified,bencini2008some,CuO2_J_Illas_00,CuO2_J_Illas_04}. There are no intrinsic complications for inclusion of effects such as spin-orbit coupling or external fields, which facilitates the calculation of a wide variety of different couplings~\cite{bastardis2007microscopic,CuO2_DM_pradipto_2012,Bogdanov13}.

The most significant disadvantage of quantum chemistry approaches lies in the much higher computational expense that scales exponentially with the size of the active space. For this reason, the full periodic crystal cannot be treated, and calculations are restricted to smaller clusters of finite size (typically two magnetic ions). In typical calculations the local crystalline environment is simulated by including nearby atoms to capture the local crystalline field and ligand effects \cite{calzado2000accurate,bencini2008some,birkenheuer2006simplified}, while additional point charges can be included to simulate the long-range electrostatic potential of the omitted atoms~\cite{bastardis2007microscopic,rev_qc_emb_carter_08}. This approach reduces as much as possible spurious finite size effects. However, calculations can become increasingly intractable for ions with larger numbers of unpaired electrons in the ground state (i.e. higher spin $S$ per site). This also limits the ability to treat longer-range or multi-site interactions, even for lower spin counts per site. In practice, quantum chemistry methods are therefore most applicable for studying first or second neighbour interactions for low-spin magnetic ions, where very reliable results may be obtained. \\


\subsection{``Hybrid'' Cluster Expansion Methods} \label{sec:3p4} In order to balance some of the advantages and disadvantages of the previous methods, the present authors have recently employed a semi-{\it ab-initio} scheme, which has been shown to yield promising results for complex spin Hamiltonians \cite{riedl2016abinitio,winter2016challenges,winter2017importance,riedl2018critical}. 
This methodology is based on the strategy of dividing the derivation of the spin Hamiltonian $\mathcal{H}_{\rm spin}$ into two steps.

First, an effective electronic Hubbard-like Hamiltonian $\mathcal{H}_{\rm eff}$ is derived, which incorporates the relevant orbital and charge degrees of freedom for the magnetic couplings. This Hamiltonian can be considered as an intermediate energy theory, where higher lying states have been integrated out, resulting in screening of the Coulomb interactions and renormalization of the kinetic energy. In contrast to quantum chemistry methods, which explicitly treat the dynamical screening, the hopping and Coulomb parameters in the intermediate energy theory are estimated from less expensive DFT-based methods, on the basis of appropriately constructed Wannier orbitals \cite{marzari1997maximally,souza2001maximally}. For example, constrained RPA \cite{miyake2008screened,PhysRevB.74.125106} may be employed to compute appropriate screened Coulomb interactions. This first step significantly reduces the size of the effective electronic Hilbert space, and the obtained intermediate model respects the full translational symmetry of the crystal.

A second procedure is then employed to obtain the low-energy Hamiltonian from the intermediate energy Hubbard model. Similar to the quantum chemistry methods, $\mathcal{H}_{\rm eff}$ is exactly diagonalized for a finite cluster of sites, in order to yield the exact low-energy eigenstates $\{|\psi_n\rangle\}$ and energies $E_n$. To map these values to an effective spin Hamiltonian, a suitable low-energy Hilbert space $\mathcal{S}_{\rm low} = \{|\phi_n\rangle\}$ of pure spin (or angular momentum) states is then selected. The eigenstates are projected into the low-energy space, following the method of des Cloizeaux~\cite{des1960extension,maurice2009universal}:
\begin{align}
|\xi_n\rangle =\mathcal{P}|\psi_n\rangle =  \sum_m |\phi_m\rangle \langle \phi_m | \psi_n\rangle
\end{align}
so that $|\xi_n\rangle \in \mathcal{S}_{\rm low}$. These intermediate states $|\xi_n\rangle$ are orthonormalized via the symmetric (L\"owdin) method~\cite{loewdin1950nonortho}, in terms of the overlap matrix $\mathbb{S}$:
\begin{align}
|\xi_n^\prime\rangle = \mathbb{S}^{-1/2} \mathcal{P} | \psi_n\rangle\\
[\mathbb{S}]_{nm} = \langle \xi_n | \xi_m\rangle
\end{align}
which defines a general and unique mapping between the intermediate and effective low-energy states. 
The effective spin Hamiltonian is then defined by the relation:
\begin{align}
\mathcal{H}_{\rm spin} |\xi^\prime_n\rangle = E_n |\xi_n^\prime\rangle
\end{align}
The basic utility of this approach is that the mapping maximizes $\langle \psi_n | \xi^\prime_n\rangle$ and preserves all symmetries~\cite{mayer2002loewdin}, which ensures that the obtained coupling constants always converge to the forms of the corresponding perturbation theory. In fact, this procedure leads to the ``optimal'' mapping discussed in Sec.~\ref{sec:concept}. This aids in the physical interpretation of the coupling constants. Furthermore, {\it all} symmetry-allowed couplings for the chosen cluster of sites are obtained simultaneously at the same level of approximation, allowing for very large and very small couplings to be estimated with similar relative accuracy. 

By sacrificing the full {\it ab-initio} quality through the use of effective intermediate electronic Hamiltonians, significantly larger numbers of sites can be treated than in typical quantum chemistry methods. This allows for some longer-range and higher order multi-site couplings to be readily estimated. Results from different cluster shapes can also be employed in order to partially mitigate errors caused by the finite truncation of the clusters. In this case, the coupling constants are interpreted as expectation values of some operator, (i.e. $J = \langle \psi| \hat{\mathcal{O}}_J|\psi\rangle$) that is amenable to calculation by a linked cluster expansion in the spirit of the Contractor Renormalization Group (CORE) method~\cite{morningstar1994contractor,morningstar1996contractor,yang2012effective}. This can be formally justified by considering the perturbative expansion for $J$ in the framework of Sec.~\ref{sec:pert}. For example, consider computing the coupling constant $J_{ij}$ for two sites $i$ and $j$ for a single orbital Hubbard model. For a two-site cluster, we obtain $J_{ij}^{\rm (2-site)} = 4t_{ij}^2/U + \mathcal{O}(t_{ij}^4/U^3)$. In contrast, for a three-site cluster, there are additional third-order contributions: $J_{ij}^{\rm (3-site)} = J_{ij}^{\rm (2-site)} + \mathcal{O}(t_{ij}t_{jk}t_{ki}/U^2)$. In order to capture such third order contributions from {\it all} 3-site clusters, we can sum over all possible third sites $k$ with little additional cost:
\begin{align}
J_{ij}^{(\infty)}= J_{ij}^{\rm (2-site)} + \sum_{k}\left(J_{ij}^{\rm (3-site)} - J_{ij}^{\rm (2-site)} \right) + ...
\end{align}
If such higher order contributions are significant (e.g.~because of $U \sim t$), then such cluster expansions may be useful for treating effectively larger clusters than computationally tractable in single calculations. Furthermore, such expansions can help to restore symmetries of the full lattice, if the only available finite clusters break some symmetries. Such ``hybrid'' methods may provide a valuable complement to more established {\it ab-initio} techniques, as exemplified by some of the applications discussed in the next section.


\section{Selected Applications to Frustrated Magnetic Materials} 

Having discussed some commonly employed numerical and analytical techniques for the derivation of low-energy spin Hamiltonians, we now turn to a number of recent case studies of frustrated magnetic materials to which some of the present authors have contributed. In each case, we wish to highlight the rational behind the application of a particular {\it ab-initio} method. \\


\subsection{Geometrically Frustrated Heisenberg Kagome Lattice MCu$_3$(OH)$_6$Cl$_2$ Family}

\begin{figure}[t]%
\includegraphics*[width=0.85\linewidth]{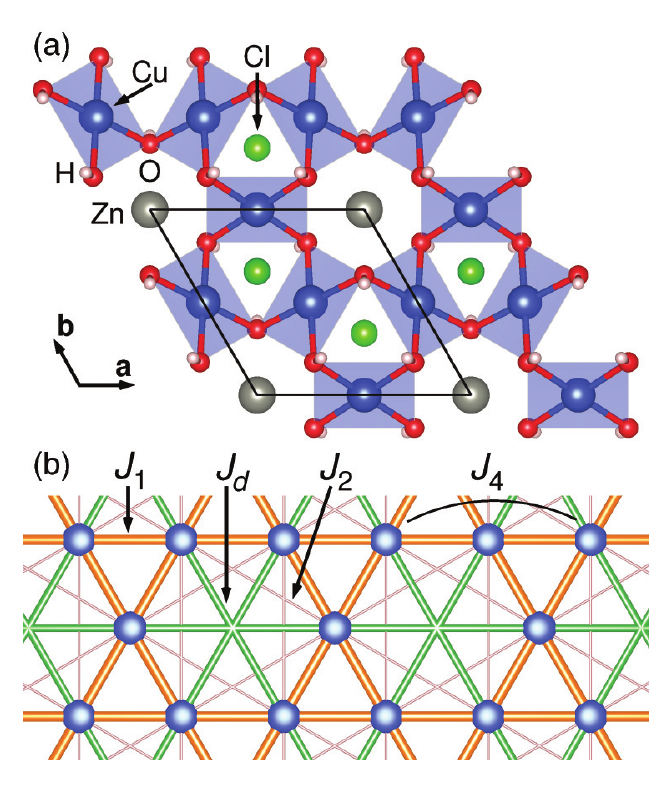}
\caption{(a) Crystal structure of kapellasite, viewed along the c direction. (b) Kagome lattice formed by the Cu sites in (a), showing unique exchange couplings. Reprinted with permission from Ref.~\onlinecite{Jeschke2013}.}
\label{fig:herb}
\end{figure}

Recently, there has been significant interest in antiferromagnetic materials adopting a two-dimensional Kagome lattice geometry featuring corner sharing triangles, such as Kappelasite shown in Fig.~\ref{fig:herb}(a). From the theoretical point of view, the ground state of the nearest neighbour quantum Heisenberg model on this lattice continues to be a subject of much interest, with leading proposals including various quantum spin liquids~\cite{PhysRevB.83.212401,PhysRevB.83.100404,yan2011spin,jiang2012identifying,PhysRevB.89.020407}.
Such theoretical studies may be complemented significantly by material realizations of this model, sparking interest in the development of Kagome materials with dominant Heisenberg couplings (i.e. weak spin-orbit coupling). 

Of those Kagome materials currently discovered, Herbertsmithite (M = Zn) is often discussed as the best realization of the antiferromagnetic $S=1/2$ Heisenberg model to date~\cite{Mendels2010,Mendels2011}. The spins are carried by Cu$^{2+}$ ions, with nearest neighbour interactions mediated by Cu$-$(OH)$-$Cu exchange pathways. Measurements of the magnetic susceptibility~\cite{Helton2007} have indicated antiferromagnetic couplings of the order $J/k_B$ $\sim$ 190 K, while no magnetic order is observed down to 50 mK~\cite{Mendels2007}. Inelastic neutron scattering experiments~\cite{Vries2009,Han2012} have also been interpreted in terms of fractionalized quantum excitations. Kagome lattice geometry is also realised by the related Kapellasite (a polymorph of Herbertsmithite with the same chemical formula), for which $\mu$SR experiments have indicated no magnetic order down to 20 mK~\cite{faak2012kapellasite}. In this case, inelastic neutron scattering reveals the development of short-range dynamical correlations consistent with a noncoplanar twelve-sublattice ``cuboc2'' magnetic structure. In contrast, the related Haydeeite~\cite{colman2010comparisons} (isostructural to Kapellasite, with M = Mg) exhibits a ferromagnetic order below $T_C \sim $ 4 K~\cite{PhysRevB.91.220408}. These contrasting experimental results motivated attempts to understand the structural dependence of the underlying Heisenberg couplings.

Within the 2D Kagome layers, the minimal effective spin Hamiltonians can be described by a sum of bilinear Heisenberg interactions~\cite{Jeschke2013,Iqbal2015}:
\begin{align}
\mathcal{H}_{\rm spin} =& \ \sum_{\langle ij \rangle} J_1 \ \mathbf{S}_i \cdot \mathbf{S}_j + \sum_{\langle \langle ij \rangle\rangle} J_2 \ \mathbf{S}_i \cdot \mathbf{S}_j \nonumber \\  
& \ + \sum_{\langle\langle\langle ij \rangle\rangle\rangle} J_d \ \mathbf{S}_i \cdot \mathbf{S}_j 
\end{align}
where $J_1$, $J_2$, and $J_d$ refer to first neighbour, second neighbour, and diagonal third neighbour couplings, as shown in Fig.~\ref{fig:herb}(b). The phase diagram of this model presents a variety of ordered and quantum disordered regions as a function of the exchange parameters \cite{Iqbal2015,PhysRevLett.108.207204,PhysRevB.89.020408,PhysRevB.92.060407}, and is particularly sensitive even towards small $J_2$ and $J_d$ couplings. 

Due to the isotropic Heisenberg nature of the interactions, and the importance of such longer-range couplings, such materials were amenable to study by broken symmetry DFT approaches. On the basis of GGA+U calculations (with $U = 6-8\,$eV for the Cu atoms), the authors of Refs.~\cite{Jeschke2013,Iqbal2015} emphasized the role of the Cu-O-Cu bonding angles in establishing the sign of $J_1$. For Herbertsmithite, the bond angle of $119^\circ$ provides a sufficiently large oxygen-mediated hopping to lead to antiferromagnetic interactions. The estimated coupling constant from broken symmetry methods was found to be $J_1 \sim +182\,$K, which is in remarkable agreement with the experimental susceptibility fits. The further neighbour couplings were estimated to be much smaller, $J_2 \sim 3.4$ K, and $J_d \sim -0.4$ K, thus justifying analogies to the simple nearest neighbour antiferromagnetic model.

In contrast, Kapellasite and Haydeeite display much smaller Cu-O-Cu bond angles of $\sim 106^\circ$ and $105^\circ$, respectively. This leads to a competition between antiferromagnetic and ferromagnetic contributions to the exchange, which enhances the importance of longer range couplings. As discussed by the authors of Ref.~\onlinecite{Iqbal2015}, such competition also requires some care in selecting the appropriate DFT functional and implementation of the double-counting corrections in the DFT+U formulation. Thus, while LDA+U calculations\cite{janson2008modified} first suggested $J_1>0$ in both materials, subsequent GGA+U calculations~\cite{Iqbal2015} indicated a significant suppression of the antiferromagnetic nearest-neighbour coupling, leading to $J_1<0$. For Kapellasite, broken symmetry GGA+U estimated that the ratio of first and third neighbour couplings was $|J_1|/J_d \sim 0.8$. For Haydeeite, it was found that $|J_1|/J_d \sim 1.6$. Crucially, the magnetic ground state is controlled by this ratio, which reflects a competition between tendencies to order as a ferromagnet for $|J_1|/J_d \gg 1$ or a cuboc-2 antiferromagnet for $|J_1|/J_d \ll 1$. Thus, the authors of Ref.~\onlinecite{Iqbal2015} were able to rationalize the differing magnetic ground states of the two structurally similar materials on the basis of microscopic details.\\


\subsection{Triangular Lattice Organics Near the Mott Transition}

\begin{figure}
\includegraphics[width=\columnwidth]{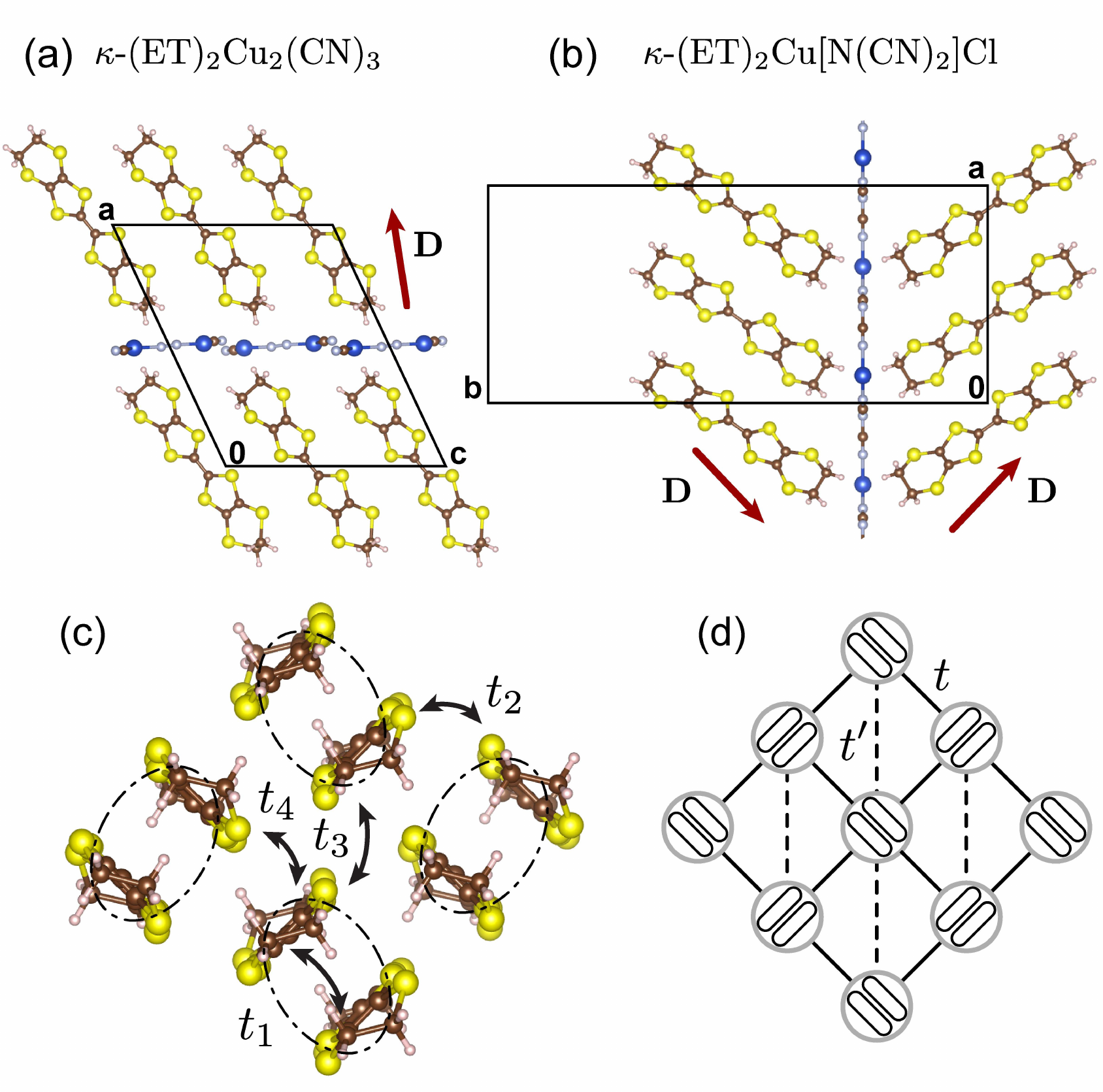}
\caption{Structure of organic materials $\kappa$-(ET)$_2$X, showing (a,b) unit cell for X = Cu$_2$(CN)$_3$ and X = Cu[N(CN)$_2$]Cl, showing the computed orientation of the DM-vector $\mathbf{D}$. (c) Unique hopping integrals within the organic layer. (d) Effective anisotropic triangular lattice of dimers. Figure adapted with permission from Ref.~\onlinecite{winter2017importance}.}
\label{fig:Organics_structure}
\end{figure}

The organic $\kappa$-(ET)$_2$X materials consist of an alternating layered
structure of organic ET dimers and inorganic counter-anions X, shown in
Fig.~\ref{fig:Organics_structure}(a,b). The inorganic layer is typically closed
shell, while each ET dimer within the organic layer has one hole, on average.
As a result, the chemical modification of the anions X allows the properties of
the organic layer to be tuned, providing examples of magnetic Mott insulators,
superconductors, and metals as the $U/t$ ratio is effectively
tuned~\cite{toyota2007low,lebed2008physics,shimizu2003spin,powell2011quantum,dressel2011quantum,lunkenheimer2012multiferroicity,tocchio2013spin,tocchio2014one,gati2016breakdown}.
In the insulating case, each hole is localized to its parent dimer by Coulomb
repulsion, occupying the anti-bonding combination of molecular HOMOs, given by
$\vert a \rangle = \frac{1}{\sqrt{2}}(\vert g_1 \rangle + \vert g_2 \rangle)$.
This gives rise to an $S=1/2$ moment per dimer. As a result, the minimal
magnetic model can be considered as a spin Hamiltonian on the anisotropic
triangular lattice, as shown in Fig.~\ref{fig:Organics_structure}(c,d).
A variety of magnetic salts have been synthesized and studied, representing different limits of the available physics on the anisotropic triangular lattice. For example, X = Cu[N(CN)$_2$]Cl orders magnetically in a square lattice N\'eel order with $T_N=27\,$K~\cite{miyagawa1995antiferromagnetic,lunkenheimer2012multiferroicity}. In contrast, there are two quantum spin liquid candidates~\cite{yamashita2008thermodynamic,shimizu2003spin,shimizu2016pressure,pinteric2016anion}, X = Cu$_2$(CN)$_3$ and X = Ag$_2$(CN)$_3$ for which a detailed determination of the corresponding spin Hamiltonian including higher order corrections is of high interest.

Various experiments have pointed to the presence of small spin-anisotropic terms \cite{smith2003dzialoshinskii,smith2004precise,kagawa2008field} arising from the effects of weak spin-orbit coupling, which have been argued to be relevant at low energies~\cite{winter2017importance,riedl2018critical}. The suppression of magnetic order in $\kappa$-Cu has further been attributed to the presence of longer range couplings and 4-spin ring-exchange interactions $K_{ijkl}$ \cite{motrunich2005variational,block2011spin,holt2014spin} due to close proximity to a delocalized metallic phase (i.e. relatively small $U/t$). Finally, the low-energy response of the spin liquid may be perturbed by magnetic field-induced 3-spin scalar chiral interactions $M_{ijk}$ \cite{motrunich2006orbital}, which have been suggested as a potential probe of a spinon Fermi surface. These interactions can be summarized by the spin-Hamiltonian:
\begin{align}
\mathcal{H}_{\rm spin} =& \  \sum_{ij} J_{ij} \ \mathbf{S}_i \cdot \mathbf{S}_j + \mathbf{D}_{ij} \cdot (\mathbf{S}_i \times \mathbf{S}_j) + \mathbf{S}_i \cdot \Gamma_{ij} \cdot \mathbf{S}_j
\nonumber
 \\ & \ + \frac{1}{S}\sum_{ijk} M_{ijk} \  \mathbf{S}_i \cdot (\mathbf{S}_j \times \mathbf{S}_k) \nonumber 
\\ & \ + \frac{1}{S^2} \sum_{ijkl} K_{ijkl} (\mathbf{S}_i\cdot\mathbf{S}_j)(\mathbf{S}_k\cdot\mathbf{S}_l)
\end{align}
where spin-orbit coupling leads to $\mathbf{D}_{ij}$, the antisymmetric Dzyalloshinskii-Moriya (DM) interaction, and $\Gamma_{ij}$, the symmetric pseudo-dipolar tensor. Estimation of all such couplings via full periodic-crystal DFT calculations have proven largely intractable, due to the delocalization of the magnetic moment across the organic molecules. The absence of localized atomic spin centers complicates the implementation of DFT+U, while hybrid DFT functionals are prohibitively expensive in band-structure codes. 

In contrast, ``hybrid'' cluster expansion methods have proved to be suitable~\cite{winter2017importance,riedl2018critical}.
In this case, the intermediate electronic Hamiltonian was taken to be $\mathcal{H}_{\rm eff} = \mathcal{H}_{\text{hop}^\ast} + \mathcal{H}_{\text{int}}$, including two orbitals per ET dimer (the highest lying bonding and antibonding orbitals) with an average filling of 3/4. The two-particle interactions were considered as a Hubbard repulsion plus Hund's coupling within each dimer. The Coulomb parameters were taken to be those estimated by constrained RPA \cite{nakamura2012abinitio}, but scaled by a factor $\approx 2/3$ after comparison to the experimental magnetic interactions. The spin-orbit coupling was projected into the extended molecular orbitals, resulting in a complex hopping term $\vec{\lambda}_{ij}$
\cite{yildrim1995anisotropic}:
\begin{align}
\mathcal{H}_{\text{hop}^\ast} = \sum_{ij} \sum_{\alpha \beta} \mathbf{c}_{i \alpha}^\dagger \left( t_{i \alpha, j \beta} \mathbbm{1}_{2 \times 2} + \frac{i}{2} \vec{\lambda}_{i \alpha, j \beta} \cdot \vec{\sigma} \right) \mathbf{c}_{j \beta}.
\end{align}
The hopping and spin-orbit parameters were estimated from DFT calculations with the quantum chemistry code ORCA~\cite{neese2012orca} employing the spin-orbit mean field approach. With the intermediate Hamiltonian established, the coupling constants were computed by cluster expansions with clusters of up to eight ET molecules (4 dimers) by projecting onto pure spin states with one hole localized to each ET dimer.

On the basis of these calculations, the authors of
Ref.~\onlinecite{winter2017importance,riedl2018critical} derived several interesting
observations related to the electronic and magnetic properties of
some representative ET-based compounds. Significant contributions from higher order terms beyond the
typical $4t^2/U$ approximation were found by studying the size convergence of
the cluster expansion. These do not affect the couplings equally, but
significantly increase the $J^\prime/J$ ratio in most compounds. Together with
ring-exchange terms found to be on the order of $K/J \sim 10\%$, this places
materials such as X = Cu$_2$(CN)$_3$ (with $J \sim $ 230 K and $J^\prime/J \sim
$ 1.2) firmly in a region expected to exhibit a spin-liquid ground state, while
X = Cu[N(CN)$_2$]Cl (with $J \sim 480$ K and $J^\prime/J \sim$ 0.3) were placed
in an ordered antiferromagnetic phase according to the phase diagrams of
Ref.~\onlinecite{holt2014spin}. In both cases, such ground states were consistent
with experimental observations. Furthermore, the weak DM-interaction
($|\mathbf{D}|/J \sim$ 5\%) of the latter compound was found to be in excellent
quantitative agreement with experimental
estimates~\cite{smith2003dzialoshinskii,smith2004precise}, in terms of both
size and orientation of $\mathbf{D}$. This finding validates the approach of
projecting the spin-orbit effects into the molecular orbital basis, and
highlights the utility of wavefunction based approaches for estimating small
anisotropic exchange couplings. \\


\subsection{Pyrochlore Ferromagnet \luvo}

\begin{figure}[t]
\includegraphics[width=\columnwidth]{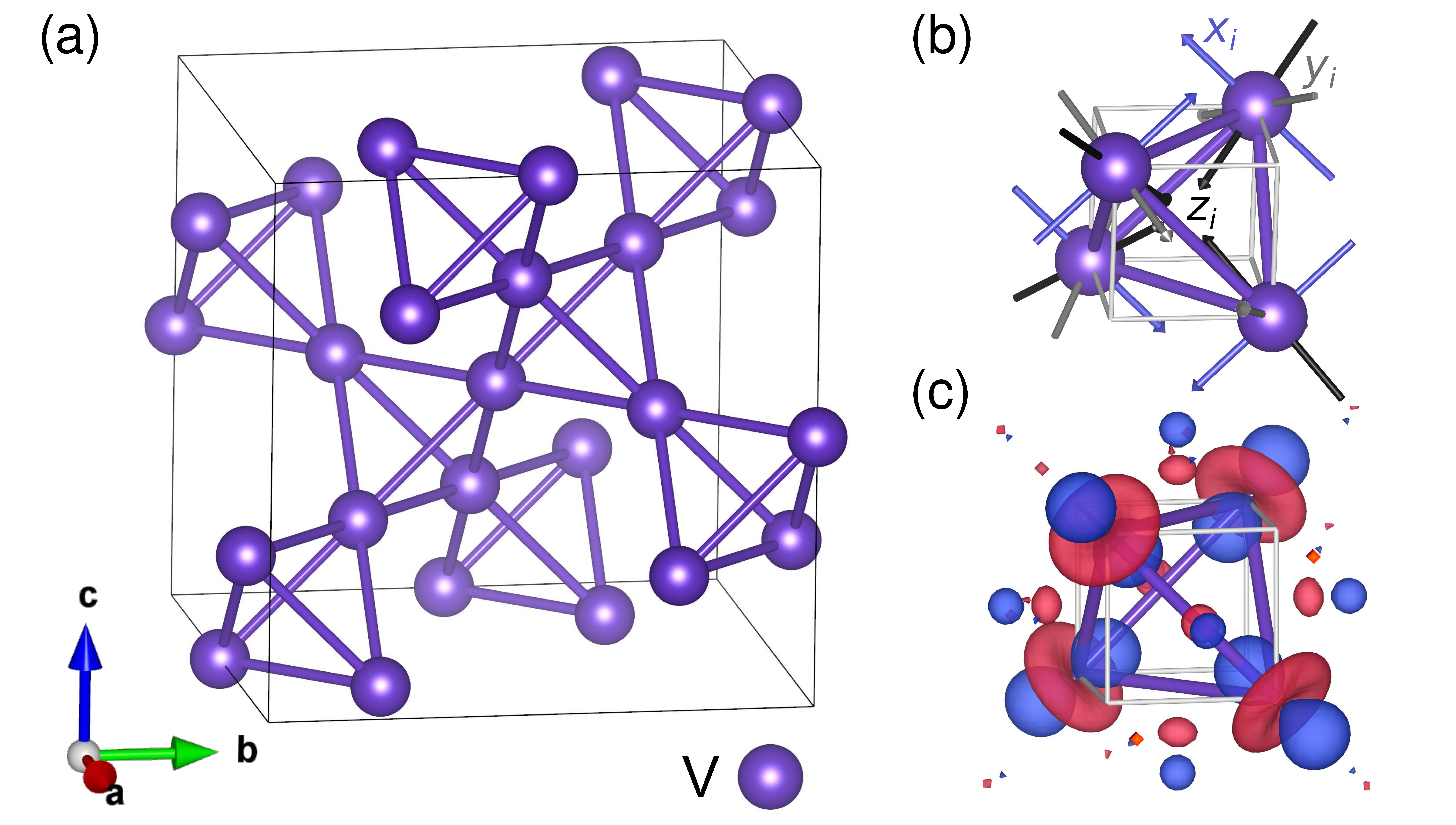}
\caption{(a) Network of corner-sharing V$^{4+}$ tetrahedra in \luvo. (b) Local coordinates in one vanadium tetrahedron. (c) Wannier $d_{z^2}$ orbitals determined with FPLO~\cite{koepernik1999fplo,eschrig2009tb} in one vanadium tetrahedron.}
\label{fig:Pyrochlore_structure}
\end{figure}

The rare-earth vanadate Lu$_2$V$_2$O$_7$ represents an interesting ferromagnetic $S = 1/2$ pyrochlore, where the magnetism is carried by a network of corner-sharing V$^{4+}$ tetrahedra with $3d^1$ configuration, shown in Fig.~\ref{fig:Pyrochlore_structure}(a). It is a Mott insulator with a Curie temperature $T_c \approx 70\,$K~\cite{shamoto2002substitution,zhou2008magnetic,onose2010observation}. Interest emerged in the anisotropic exchange interactions in \luvo, after the observation of a large magnon Hall effect \cite{onose2010observation}. It was argued that a finite Dzyaloshinskii-Moriya interaction plays the role of a vector potential in the electronic case. The effect of a finite DM interaction in a ferromagnetic pyrochlore structure is thought to induce a topological magnon insulator state with chiral edge modes~\cite{mook2014magnon,zhang2013topological} and the appearance of magnon Weyl points~\cite{mook2016tunable}.

From an experimental standpoint, the underlying spin Hamiltonian remains somewhat controversial. From fits of the magnetic specific heat and of the transverse thermal conductivity, the authors of Ref.~\onlinecite{onose2010observation} estimated $\vert \mathbf{D}_{ij} / J_{ij} \vert = 0.32$ with a nearest neighbour Heisenberg interaction $J_{ij} \approx -3.4$~meV. However, in Ref.~\onlinecite{mook2014magnon} it was argued that refitting the data with additional corrections suggested a ratio two orders of magnitude smaller with $\vert \mathbf{D}_{ij}/J_{ij} \vert \approx 0.005$. Inelastic neutron scattering~\cite{mena2014spinwave} fitting reveiled a larger ferromagnetic Heisenberg exchange than the transport data, with $J_{ij}=-8.1\,$meV and a somewhat smaller ratio $\vert \mathbf{D}_{ij} / J_{ij} \vert \approx 0.18$ from fittings in specific regions in reciprocal space. 

From a microscopic perspective, the origin of the ferromagnetic sign of $J$ was initially discussed in detail in Ref.~\onlinecite{shamoto2002substitution,ichikawa2005orbital,miyahara2007orbital}. The V$^{4+}$ ions are in a trigonally distorted octahedral environment, which leads to the single electron occupying a Wannier orbital of $d_{z^2}$ symmetry with respect to the local coordinates shown in Fig.~\ref{fig:Pyrochlore_structure}(b,c). Due to the nearly empty $d$-shell, a large number of excited triplet configurations exist with two electrons at the same V site, which are stabilized by Hund's coupling $J_H$ and mixed into the neutral ground state by significant inter-orbital hopping. Thus, it was expected that the $J<0$ arose from a subtle competition between typical antiferromagnetic contributions $\sim (t_{z^2\to z^2})^2/U$ and ferromagnetic contributions like $\sim (t_{z^2 \to xy})^2J_H/U^2$. This idea was justified by analytical perturbation expressions \cite{arakawa2016microscopic}, although such calculations included only contributions from the lowest three of the $d$-orbitals and could not address the ratio of $|\mathbf{D}_{ij}/J_{ij}|$ without {\it ab-initio} parameters.

From the {\it ab-initio} perspective, DFT estimates based on non-collinear spin configurations were presented in Ref.~\onlinecite{xiang2011singleion}, which estimated $J_{ij} = -7.1$ meV and $|\mathbf{D}_{ij}/J_{ij}| \approx 0.05$. However, the authors also noted a significant single ion anisotropy contribution in the DFT configuration energies, which would be forbidden in the true quantum Hamiltonian for $S = 1/2$ spins. As a result, the reliability of the effective classical mapping was not clear. For this reason, the authors of Ref.~\onlinecite{riedl2016abinitio} explored the use of wavefunction-based ``hybrid'' methods, which would capture such restrictions imposed by the quantization of the spin.

\begin{figure}[t]
\includegraphics[width=0.95\columnwidth]{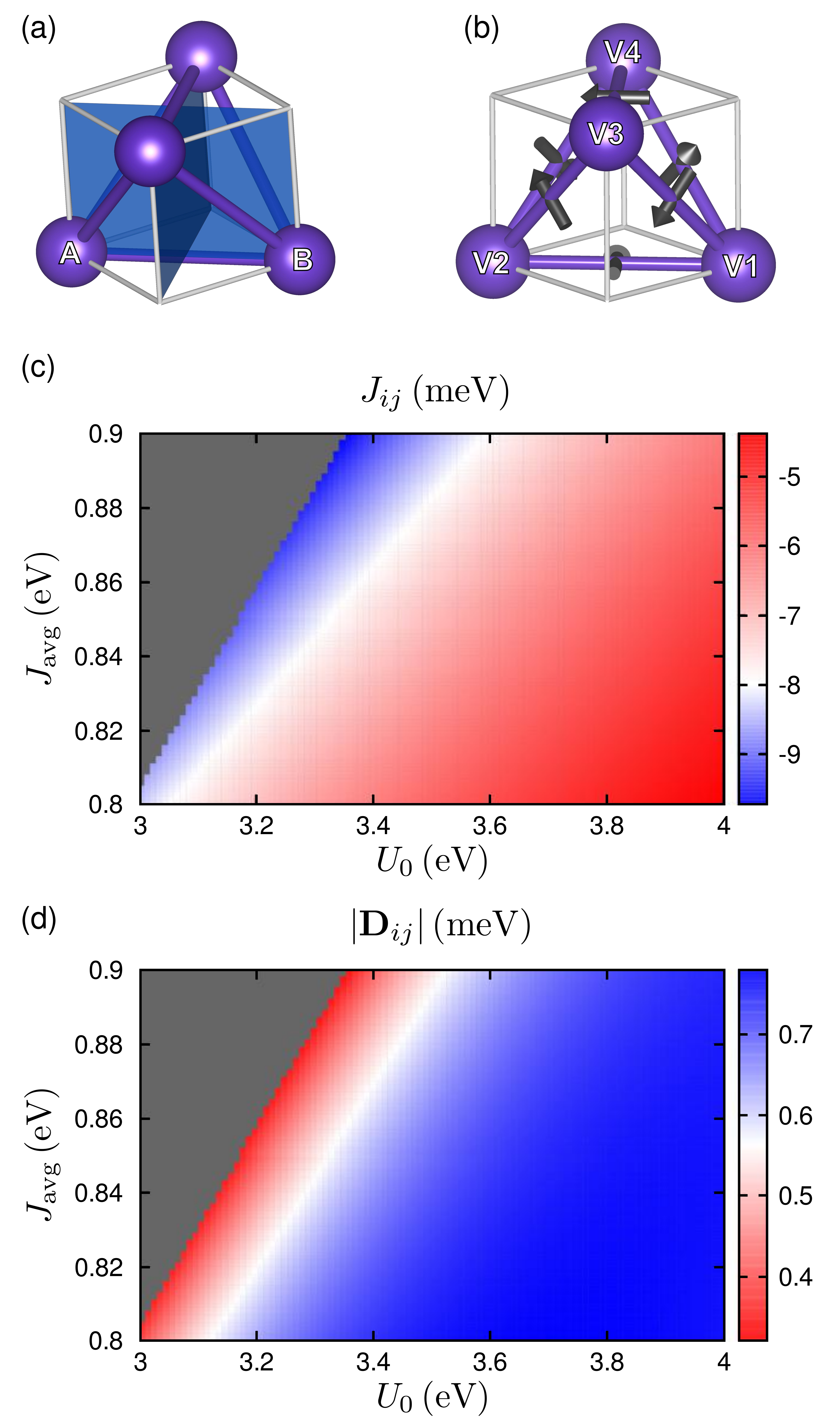}
\caption{(a) Two mirror planes with respect to the magnetic sites A and B in the pyrochlore lattice. (b) DM vectors in the pyrochlore lattice, with sign choice corresponding to the ``indirect'' case defined in Ref.~\onlinecite{elhajal2005ordering}.  (c,d) Computed couplings from Ref.~\onlinecite{riedl2016abinitio}. Unphysical regions with $U_0-3J_{\alpha \beta}<0$ are indicated in grey. Figure adapted with permission from Ref.~\onlinecite{riedl2016abinitio}.}
\label{fig:Pyrochlores_symmetry}
\end{figure}

In order to estimate the couplings, the authors of Ref.~\onlinecite{riedl2016abinitio} considered the $d$-orbital Wannier orbitals, as described by the intermediate electronic Hamiltonian $\mathcal{H}_{\text{tot}} = \mathcal{H}_{\text{hop}} + \mathcal{H}_{\text{SOC}} + \mathcal{H}_{\text{int}}$. Nearest neighbour hopping parameters were determined via the projective Wannier functions as implemented in the all-electron full-potential local orbital code FPLO~\cite{koepernik1999fplo,eschrig2009tb}. The spin-orbit coupling operator within this basis takes the general form:
\begin{align}
\mathcal{H}_{\text{SOC}} = \lambda \sum_i \sum_{\alpha \beta} \sum_{\sigma \sigma^\prime} \langle i \alpha \sigma \vert \mathbf{L} \cdot \mathbf{S} \vert i \beta \sigma^\prime \rangle d_{i \alpha \sigma}^\dagger d_{j \beta \sigma^\prime},
\end{align}
while the Coulomb interactions are:
\begin{align}
\mathcal{H}_{\text{int}} 
=& \sum_i \sum_{\alpha \beta} U_{\alpha \beta} n_{i \alpha \uparrow} n_{j \beta \downarrow} \nonumber \\
&+ \frac{1}{2} \sum_{i \sigma} \sum_{\alpha \neq \beta} (U_{\alpha \beta} - J_{\alpha \beta} ) n_{i \alpha \sigma} n_{j \beta \sigma^\prime} \nonumber \\
&+ \sum_i \sum_{\alpha \neq \beta} J_{\alpha \beta} \nonumber
( d_{i \alpha \uparrow}^\dagger  d_{i \beta \downarrow}^\dagger  d_{i \alpha \downarrow}  d_{i \beta \uparrow} \\ & 
\hspace{20mm}+ d_{i \alpha \uparrow}^\dagger  d_{i \alpha \downarrow}^\dagger  d_{i \beta \downarrow}  d_{i \beta \uparrow}),
\end{align}
The authors suggested that the matrix elements of these operators in the Wannier orbital basis could be well approximated by their action on the equivalent $d$-orbitals of the free ion. Thus, the Coulomb terms were approximated  by two independent parameters~\cite{liechtenstein1995density,pavarini2011dmft}: the Coulomb repulsion of electrons on the same site and orbital $U_0$ and the average Hund's coupling $J_{\text{avg}}=\frac{1}{2l(2l+1)} \sum_{\alpha \neq \beta} J_{\alpha \beta}$. For the spin-orbit coupling, fitting the difference between band structures at the DFT and DFT+SOC within this approximation 
yielded a spin-orbit constant of $\lambda = 30.0\,\text{meV}$, which is very close to the reported free ion value of $\lambda_{\text{exp}}= 30.75\,\text{meV}$~\cite{blume1963theory}. Thus, interestingly, the spin-orbit coupling was not found to be strongly renormalized by projection into the Wannier basis.

Various schemes have been used in the literature to parameterize the symmetry-allowed interactions for the pyrochlore structure~\cite{thompson2011rods,ross2011quantum,riedl2016abinitio}, which include antisymmetric DM-interactions, as well as symmetric anisotropic exchange. The authours of Ref.~\onlinecite{riedl2016abinitio} followed the general scheme:
\begin{align}
 \mathcal{H}_{\rm spin} = \sum_{ij} J_{ij} \mathbf{S}_i \cdot \mathbf{S}_j + \mathbf{D}_{ij} \cdot (\mathbf{S}_i \times \mathbf{S}_j) + \mathbf{S}_i \cdot \Gamma_{ij} \cdot \mathbf{S}_j.
 \end{align}
 
Mirror symmetry (Fig.~\ref{fig:Pyrochlores_symmetry}(a)) constrains the
DM-vectors to point along specific axes perpendicular to each bond, as shown in
Fig.~\ref{fig:Pyrochlores_symmetry}(b). In order to estimate these couplings,
the authors of Ref.~\onlinecite{riedl2016abinitio} projected the lowest energy states
of the 2-site model onto pure spin states with one electron per $d_{z^2}$
orbital, and studied the resulting couplings as a function of the screened
Coulomb terms $U_0$ and $J_{\rm avg}$. The Heisenberg and DM-interactions are
summarized in Fig.~\ref{fig:Pyrochlores_symmetry}(c,d). Several conclusions
were drawn: (i) By including all five $d$-orbitals explicitly, the nearest
neighbour Heisenberg exchange $J_{ij}$ was found to be ferromagnetic in the
whole range of reasonable parameters, with a range of magnitudes in agreement
with previous experimental and DFT estimates. (ii) Interestingly, the
magnitudes of $J_{ij}$ and $\mathbf{D}_{ij}$ show opposite trends with $U_0$
and $J_{\rm avg}$, reflecting a different microscopic origin
\cite{arakawa2016microscopic}. Thus, for a value of the Heisenberg exchange
consistent with inelastic neutron scattering, the hybrid method estimated a
wide range of DM-interaction strengths $\vert \mathbf{D}_{ij}/J_{ij}
\vert=0.04-0.09$. Such magnitudes validated the previous estimates from DFT,
suggesting possible overestimation of the experimentally reported ratios. (iii)
The ability to estimate {\it all} exchange constants revealed possibly relevant
contributions from the previously ignored pseudo-dipolar tensor with $\vert
\vert \Gamma_{ij} \vert \vert / \vert J_{ij} \vert=0.01-0.02$, being not that
far from the order of magnitude of the DM interaction.


\subsection{Kitaev Magnets} 

\begin{figure}
\includegraphics[width=0.9\columnwidth]{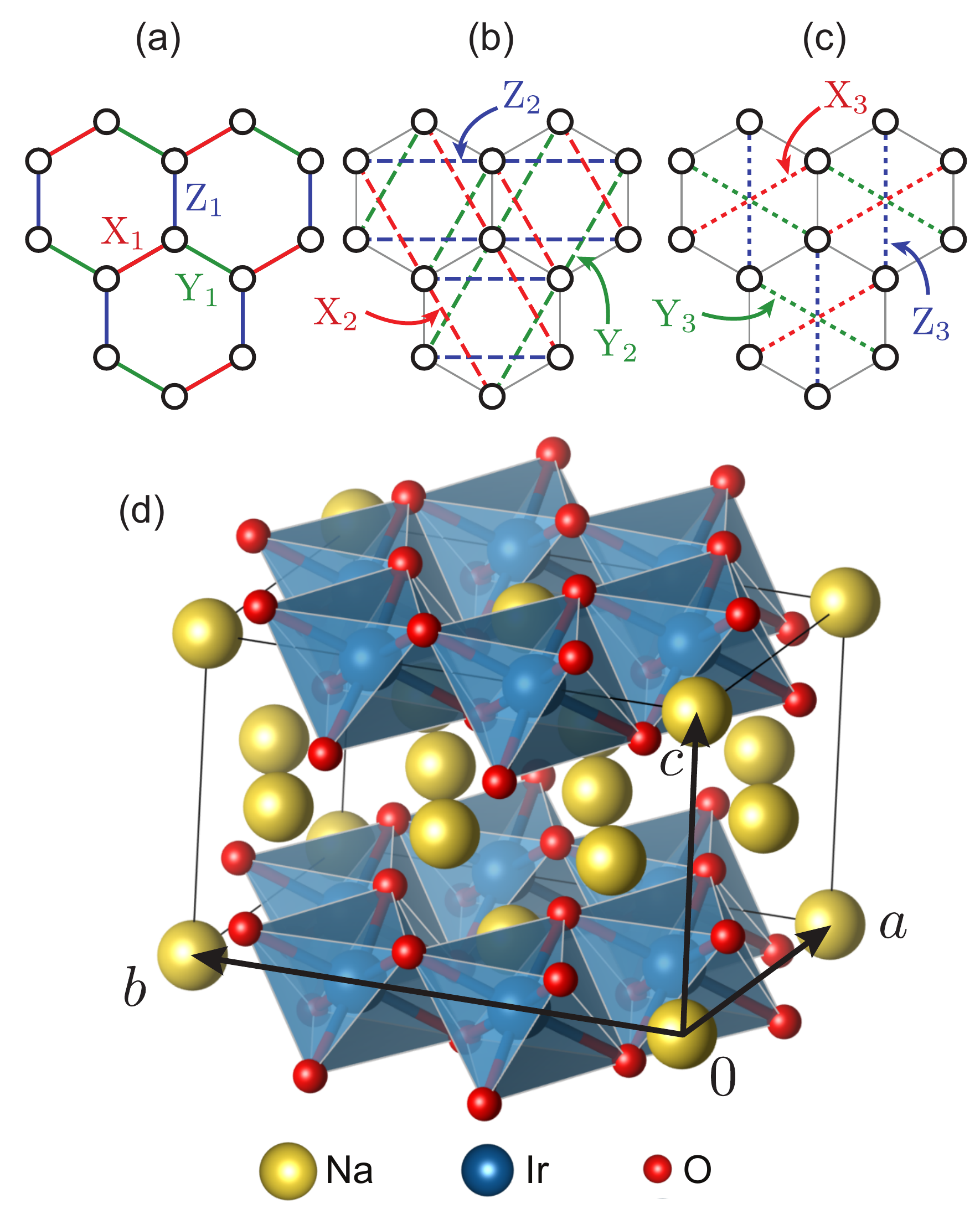}
\caption{(a-c) Definition of the first, second, and third neighbour bonds in honeycomb Kitaev materials, respectively.  (d) Structure of Na$_2$IrO$_3$. Figure (a-c) adapted with permission from Ref.~\onlinecite{winter2016challenges}.}
\label{fig:na2iro3}
\end{figure}

Recently, great interest has developed towards potential experimental realizations of Kitaev's $S=1/2$ honeycomb model~\cite{kitaev2006anyons}, which is exactly solvable and yields
a $Z_2$  spin liquid ground state. Within this model, the interactions are strongly anisotropic Ising couplings, described by the Hamiltonian $\mathcal{H} = K_1\sum_{ij} S_i^\gamma S_j^\gamma$, where the $\gamma$ axes are defined for each nearest neighbour bond, following the pattern in Fig.~\ref{fig:na2iro3}(a). 

\begin{table*}[t]
\caption{Coupling constants for the ``Z-bond'' of Na$_2$IrO$_3$, estimated via Broken Symmetry DFT (BS-DFT) \cite{hu2015first}, Quantum Chemistry (QC) \cite{katukuri2014kitaev}, and hybrid cluster methods \cite{winter2016challenges}. The BS-DFT values reflect the variable moment, all-data fitting scheme in supplementary table S3 of Ref.~\onlinecite{hu2015first}.\label{tab:one}}
\centering\def\arraystretch{1.1}
\label{tab:mag}
\begin{tabular}{l|rrrr|rrrrc|rrrr}
Method &$J_1$ & $K_1$ &$\Gamma_1$ &$\Gamma_1^{\prime}$  & $J_2$ & $K_2$ & $\Gamma_2$ & $\Gamma_2^\prime$ &$\mathbf{D}_2$ &$J_3$ & $K_3$ &$\Gamma_3$ &$\Gamma_3^{\prime}$ \\
\hline
BS-DFT &+7.2&-38.2&+1.5&-3.5&-1.6&$-$&$-$&$-$&$-$&+7.8&$-$&$-$&$-$  \\
QC Cluster (2-site)&+5.0&-20.5&+0.5&$-$&$-$&$-$&$-$&$-$&$-$&$-$&$-$&$-$&$-$ \\
``Hybrid'' (6-site)&+1.6&-17.9&-0.1&-1.8&+0.1&-1.2&+0.6&-0.3&-(0.2, 0.2, 0.1)&+6.8&+0.3&-0.2&-0.1
\end{tabular}
\end{table*}

On the basis of perturbation theory, the authors of Ref. \onlinecite{Jackeli2009} noted that such interactions could be realized, in principle, by heavy metal oxides or halides featuring edge-sharing MX$_6$ octahedra, where X = e.g. O, Cl, and M is a metal with a low-spin $d^5$ electronic configuration. In this case, strong spin-orbit coupling splits the $t_{2g}$ states into multiplets with effective angular momentum $j_{\rm eff} = 3/2$ and 1/2. For a $d^5$ configuration, one electron occupies the $j_{\rm eff} = 1/2$ state, giving rise to a local pseudo-spin degree of freedom. Written in terms of such pseudo-spins ``$\mathbf{S}_i$'', the lowest order couplings take the Kitaev form with ferromagnetic $K_1 < 0$, due to subtle effects of Hund's coupling between excited multiplets~\cite{winter2017models,PhysRevLett.112.077204,winter2016challenges}. With this observation, an explosive interest began in synthesizing and studying such materials. This has led to various studies of materials such as the iridates Na$_2$IrO$_3$~\cite{singh2012,chun2015} and various phases of Li$_2$IrO$_3$~\cite{singh2012,williams2016,takayama2015,modic2014}, as well as
 $\alpha$-RuCl$_3$~\cite{PhysRevB.91.241110,PhysRevB.91.144420,johnson2015monoclinic,PhysRevLett.114.147201,banerjee2016proximate}. While progress in this field has been reviewed
 elsewhere~\cite{winter2017models,rau2016spin,trebst2017kitaev,hermanns2018physics}, we focus here briefly on the application of {\it ab-initio} methods to the understanding of the magnetic couplings.

In these materials, deviations from the ideal scenario lead to additional interactions, including couplings beyond first neighbour. In general, the interactions can be written:
\begin{align}
\mathcal{H}_{\rm spin} = \sum_{ij} \mathbf{S}_i \cdot \mathbb{J}_{ij} \cdot \mathbf{S}_j
\end{align}
For the 2D honeycomb systems Na$_2$IrO$_3$, $\alpha$-Li$_2$IrO$_3$ and $\alpha$-RuCl$_3$, the first ($n=1$) and third ($n=3$) neighbour interactions shown in Fig.~\ref{fig:na2iro3}(a,c) approximately exhibit $C_{2h}$ symmetry, so that the interactions can be parameterized by four constants. The interaction tensors along the X-, Y-, and Z-bonds are:
\begin{align}
\mathbb{J}_{n}^{\rm X} =& \  \left(\begin{array}{ccc}J_n + K_n & \Gamma_n^\prime &  \Gamma_n^\prime \\  \Gamma_n^\prime & J_n & \Gamma_n \\ \Gamma_n^\prime  & \Gamma_n & J_n \end{array} \right) \ , \
 \mathbb{J}_{n}^{\rm Y} = \left(\begin{array}{ccc}J_n & \Gamma_n^\prime &  \Gamma_n \\  \Gamma_n^\prime & J_n+K_n & \Gamma_n^\prime \\ \Gamma_n & \Gamma_n^\prime & J_n \end{array} \right) \nonumber \\
\mathbb{J}_{n}^{\rm Z} =& \  \left(\begin{array}{ccc}J_n & \Gamma_n&  \Gamma_n^\prime \\  \Gamma_n& J_n & \Gamma_n^\prime \\ \Gamma_n^\prime  & \Gamma_n^\prime & J_n +K_n \end{array} \right)
\end{align}
For second neighbour bonds (Fig.~\ref{fig:na2iro3}(b)), the lower symmetry
allows also a finite Dzyaloshinskii-Moriya interaction $\mathbf{D}_2 \cdot
(\mathbf{S}_i \times \mathbf{S}_j)$, which has been suggested \cite{winter2016challenges}
 to play a role in
establishing an incommensurate spiral magnetic order in $\alpha$-Li$_2$IrO$_3$
\cite{williams2016}. Given the large number of parameters
in the Hamiltonian, extracting them all from experiment without guidance from
{\it ab-initio} calculations presents a formidable challenge. As a result, {\it
ab-initio} studies have played a prominent role in the development of the
field. 

Here, we focus on the interesting case study of the honeycomb iridate Na$_2$IrO$_3$, which exhibits an antiferromagnetically ordered ground state with zigzag configurations~\cite{choi2012,ye2012}, rather than the desired spin liquid. This magnetic order was unexpected \cite{PhysRevLett.110.097204}, as the zigzag state is more stable for antiferromagnetic $K_1 > 0$ \cite{PhysRevLett.112.077204}. This led the authors of Ref.~\onlinecite{PhysRevLett.110.097204} to question the validity of the original perturbative results, and discussed additional terms if the higher lying $e_g$ or ligand orbitals were considered. Instead, {\it ab-initio} studies have largely validated the original perturbative results.

On the basis of perturbation theory expressions and DFT hopping integrals, the authors of Ref.~\onlinecite{Foyevtsova2013} first noted that the nearest neighbour Kitaev coupling $K_1$ was likely to be ferromagnetic for the relevant parameter regime. Subsequently, Ref.~\onlinecite{katukuri2014kitaev} used quantum chemistry approaches to estimate the nearest neighbour $J_1,K_1$, and $\Gamma_1$ terms, confirming a dominant ferromagnetic $K_1 < 0$, as shown in Table \ref{tab:one}. This suggested the additional contributions discussed by Ref.~\onlinecite{PhysRevLett.110.097204} were not sufficient to reverse the sign of the coupling. Since the derived nearest-neighbour Hamiltonian failed to reproduce the zigzag order, the authors speculated on the existence of longer range couplings such as $J_2$ and $J_3$. These were not possible to estimate by quantum chemistry techniques due to high computational expense of including more than two magnetic sites. 
On this basis, broken symmetry DFT approaches were employed in Ref.~\onlinecite{hu2015first}. The authors of Ref.~ \onlinecite{hu2015first} discussed in detail various schemes for fitting the DFT energies to an effective spin Hamiltonian. In particular, they noted significant differences between Hamiltonians estimated by assuming a fixed moment, or accounting for the converged magnetic moments of each configuration $-$ as well as including or excluding some higher energy configurations in the fitting. The authors also considered a simplified model including only $J_2$ and $J_3$ long-range interactions. Nonetheless, all such schemes provided similar conclusions, confirming the speculations of Ref.~\onlinecite{katukuri2014kitaev}: the largest nearest neighbour couplings are ferromagnetic $K_1$, with large antiferromagnetic third neighbour $J_3$ stabilizing the observed zigzag order. 

Subsequently, ``hybrid'' methods were employed in Ref.~\onlinecite{winter2016challenges} in an attempt to estimate {\it all} coupling constants up to third neighbour. 
For such studies, the authors constrained the effective electronic Hamiltonian to the $t_{2g}$ Wannier orbitals at each Ir site, which was justified by the previous quantum chemistry results \cite{katukuri2014kitaev}. The low-energy Hilbert space was constructed by projecting on to ideal $j_{\rm eff}$ states, and clusters of up to six sites were considered. The derived couplings were found to be in quantitative agreement with the other methods, as shown in Table \ref{tab:one}. By studying the scaling of the interactions with different parameters, it was noted that {\it all} second-neighbour couplings were likely to be small due to subtle cancellations of competing terms, while the third neighbour Heisenberg coupling was particularly enhanced by higher order hopping processes. The derived parameters were found to be consistent with the zigzag order, and orientation of the ordered moments as probed by inelastic x-ray scattering~\cite{chun2015}. These observations further cemented the conclusions of the previous {\it ab-initio} works, and justified the truncated couplings considered in the broken symmetry DFT fitting. They also emphasize the complementary nature of various {\it ab-initio} methods for studying such complex magnetic materials.


\section{Outlook} 

In this work, we have reviewed some of the basic ideas and methods for the {\it ab-initio} construction of low-energy spin Hamiltonians. Such methods have been strongly influential in the study of frustrated magnetic materials in particular, as the magnetic response and ground states are often particularly sensitive to small details of the spin couplings. 

Most recently, there has been increasing interest in studying systems where spin-orbit coupling induces additional anisotropic couplings such as the antisymmetric Dzyaloshinskii-Moriya coupling. Such anisotropic interactions could, in principle, be exploited to yield interesting states such as topological magnon insulators or quantum spin-orbital liquids such as found in Kitaev's honeycomb model. The realization of such theoretical spin Hamiltonians in real materials is often initially motivated by insights from analytical perturbation theory focusing on idealized cases. This allows interesting parameter regimes to be transparently selected for further study. As materials approximating such regimes begin to be synthesized, understanding their experimental properties requires considering departures from the ideal models that inspired their study. In particular, it becomes useful to identify how such departures affect their low-energy Hamiltonians, and how to tune the couplings via chemical or physical means. This is where {\it ab-initio} studies become essential as a tool to be used in conjunction with other experimental and theoretical approaches.

With this outlook, we look forward to future developments in different {\it ab-initio} methods, which offer complementary advantages. At present, a pressing issue is the apparent outstanding challenges toward the general application of non-collinear DFT+U methods towards the estimation anisotropic exchange couplings \cite{0953-8984-30-27-275802} in solids. For wavefunction-based quantum chemistry methods, the main challenges are related to reducing computational expense, in order to treat larger systems on par with DFT. In this regard, semi-{\it ab-initio} ``hybrid'' methods appear to offer promise, by combining attractive aspects of the various methods to yield sufficient accuracy to interpret experiments. In conjunction, approximate configuration interaction solvers based on DMRG and FCIQMC \cite{sharma2012spin,booth2009fermion,cleland2011study,li2016combining,bogdanov2018new} seem to be particularly promising for increasing the size of tractable active spaces. As discussed in Sec.~\ref{sec:3p4}, wavefunction-based calculations allow, in principle, a systematic and unique derivation of the low-energy Hamiltonian of {\it any} material, provided sufficiently local effective degrees of freedom. Although this short review has focused on spin Hamiltonians, access to larger active spaces also allows for more complex cases of e.g. spin, charge, and orbital entanglement to be studied with ease. In this regard, the power and utility of {\it ab-initio} methods toward the development of functional materials appears to be ever expanding.

\section{Acknowledgments}

The authors thank the Deutsche Forschungsgemeinschaft for funding through Transregio TRR 49.

\bibliography{SFB}

\end{document}